\documentclass[12pt,thmsa]{article}
\usepackage{sw20lart}

\input tcilatex
\QQQ{Language}{
British English
}

\begin{document}

\author{B.V.Ivanov \\
Institute for Nuclear Research and Nuclear Energy, Bulgarian Academy of\\
Sciences, Tzarigradsko Shausse 72, Sofia 1784, Bulgaria}
\title{The importance of anisotropy for relativistic fluids with spherical symmetry}
\maketitle

\begin{abstract}
It is shown that an effective anisotropic spherically symmetric fluid model
with heat flow can absorb the addition to a perfect fluid of pressure
anisotropy, heat flow, bulk and shear viscosity, electric field and null
fluid. In most cases the induction of effective heat flow can be avoided.
\end{abstract}

\section{Introduction}

Spherically symmetric perfect fluid solutions in general relativity have
been studied from its very beginning, starting with the interior
Schwarzschild solution. Gradually different mechanisms in stellar models
have been identified that create pressure anisotropy \cite{one}, \cite{two}.
Radiating spherical collapse demands the introduction of heat flow \cite
{three} or null fluid which describe energy dissipation in different
approximations \cite{four}. More realistic fluids possess also bulk and
shear viscosity \cite{five}, \cite{six}, \cite{seven}. Charged perfect
fluids or dust have been discussed by many authors \cite{eight}. It has been
shown too that the sum of two perfect fluids, two null fluids or a perfect
and a null fluid can be represented by effective anisotropic fluid models 
\cite{nine}. In view of the many existing relations among the fluid models
mentioned above it is hard to point out some centre.

In this paper it is shown that the anisotropic fluid with heat flow is in
some sense the most fundamental model and can absorb the addition of
viscosity, charge and null fluids. Something more, heat flow is not
generated in most cases.

In Sec.2 the basic anisotropic fluid model is defined. In Sec.3 the fluid is
supplied with bulk and shear viscosity which leads to a new effective
anisotropic model. In Sec.4 the same is done for the addition of charge and
in Sec.5 null fluids are accommodated into the anisotropic model. Sec.6
summarizes all additions and the effective characteristics of the
fundamental anisotropic model are given. Several conclusions are drawn. In
Sec 7 the static case is discussed and a formula for the general solution is
presented. Sec 8 contains some basic conclusions.

\section{Anisotropic fluid model with heat flow}

Einstein's field equations are given by 
\begin{equation}
8\pi T_{\alpha \beta }=G_{\alpha \beta }  \label{one}
\end{equation}
where $G_{\alpha \beta }$ is the Einstein tensor, $T_{\alpha \beta }$ is the
energy-momentum tensor (EMT) and units are used so that $c=G=1$. The general
spherically symmetric metric is written as 
\begin{equation}
ds^2=-A^2dt^2+B^2dr^2+R^2\left( d\theta ^2+\sin ^2\theta \ d\varphi ^2\right)
\label{two}
\end{equation}
where $A,B,R$ are positive functions of $t$ and $r$ only. The spherical
coordinates are numbered as $x^0=t,x^1=r,x^2=\theta ,x^3=\varphi $. The
Einstein tensor involves the Ricci tensor and scalar which are given by the
metric and its first and second derivatives \cite{four}. Its non-trivial
components are $G_{00},G_{01},G_{11},G_{22}=\sin ^{-2}\theta \ G_{33}.$

We are interested in the structure of EMT. It reads for anisotropic fluids
with heat flow 
\begin{equation}
T_{\alpha \beta }=\left( \mu +p_t\right) u_\alpha u_\beta +p_tg_{\alpha
\beta }+\left( p_r-p_t\right) \chi _\alpha \chi _\beta +q_\alpha u_\beta
+u_\alpha q_\beta .  \label{three}
\end{equation}
Here $\mu $ is the energy density, $p_r$ is the radial pressure, $p_t$ is
the tangential pressure, $u^\alpha $ is the four velocity of the fluid (a
timelike vector), $\chi ^\alpha $ is a unit spacelike vector along the
radial direction and $q^\alpha $ is the heat flux (in the radial direction
too). We have 
\begin{equation}
u^\alpha u_\alpha =-1,\quad \chi ^\alpha \chi _\alpha =1,\quad u^\alpha \chi
_\alpha =0,\quad u^\alpha q_\alpha =0.  \label{four}
\end{equation}
It is assumed that the coordinates are comoving, hence, the fluid is
motionless in them 
\begin{equation}
u^\alpha =A^{-1}\delta _0^\alpha ,\quad \chi ^\alpha =B^{-1}\delta _1^\alpha
,\quad q^\alpha =qB^{-1}\delta _1^\alpha   \label{five}
\end{equation}
where $q=q\left( r,t\right) $. This gives 
\begin{equation}
T_{00}=\mu A^2,\quad T_{01}=-qAB,\quad T_{11}=p_rB^2,\quad T_{22}=p_tR^2,
\label{six}
\end{equation}
which should be plugged in the Einstein equations. They become 
\begin{equation}
8\pi \mu =A^{-2}\left( 2\frac{\dot B}B+\frac{\dot R}R\right) \frac{\dot R}%
R-B^{-2}\left[ 2\frac{R^{\prime \prime }}R+\left( \frac{R^{\prime }}R\right)
^2-2\frac{B^{\prime }R^{\prime }}{BR}-\frac{B^2}{R^2}\right] ,  \label{seven}
\end{equation}
\begin{equation}
8\pi p_r=-A^{-2}\left[ 2\frac{\ddot R}R-\left( 2\frac{\dot A}A-\frac{\dot R}%
R\right) \frac{\dot R}R\right] +B^{-2}\left( 2\frac{A^{\prime }}A+\frac{%
R^{\prime }}R\right) \frac{R^{\prime }}R-\frac 1{R^2},  \label{eight}
\end{equation}
\begin{eqnarray}
8\pi p_t &=&-A^{-2}\left[ \frac{\ddot B}B+\frac{\ddot R}R-\frac{\dot A}%
A\left( \frac{\dot B}B+\frac{\dot R}R\right) +\frac{\dot B\dot R}{BR}\right]
+  \label{nine} \\
&&\ B^{-2}\left[ \frac{A^{\prime \prime }}A+\frac{R^{\prime \prime }}R-\frac{%
A^{\prime }B^{\prime }}{AB}+\left( \frac{A^{\prime }}A-\frac{B^{\prime }}%
B\right) \frac{R^{\prime }}R\right] ,  \nonumber
\end{eqnarray}
\begin{equation}
4\pi qAB=\frac{\dot R^{\prime }}R-\frac{\dot BR^{\prime }}{BR}-\frac{%
A^{\prime }\dot R}{AR}.  \label{ten}
\end{equation}
Here the dot means time derivative and the prime is a radial derivative.

The anisotropic fluid does not radiate when $q=0$ and becomes perfect when $%
p_r=p_t$. Thus it accommodates anisotropy of pressure and heat flow when one
starts with a perfect fluid model.

Let us see now what happens when other EMTs are added to the basic
anisotropic one.

\section{Bulk and shear viscosity}

Bulk viscosity \cite{five}, \cite{six}, \cite{seven} adds to the basic EMT
the following piece 
\begin{equation}
T_{\alpha \beta }^B=-\zeta \Theta h_{\alpha \beta }  \label{eleven}
\end{equation}
where $\zeta $ is a coefficient, $\Theta $ is the expansion of the fluid and 
$h_{\alpha \beta }$ is the projector on the hyperplane orthogonal to $%
u^\alpha $%
\begin{equation}
\Theta =u_{\;;\alpha }^\alpha ,\quad h_{\alpha \beta }=g_{\alpha \beta
}+u_\alpha u_\beta .  \label{twelve}
\end{equation}
Obviously, this means the appearance of effective pressures 
\begin{equation}
p_r^B=p_t^B=-\zeta \Theta ,  \label{thirteen}
\end{equation}
which should be added to $p_r,p_t$. They do not change the degree of
anisotropy $\Delta p=p_r-p_t$. Thus even perfect fluid can absorb the bulk
viscosity. The quantities $\mu ,q$ remain the same. No heat flow is
generated in particular.

Shear viscosity is responsible for the piece 
\begin{equation}
T_{\alpha \beta }^S=-2\eta h_{\alpha \gamma }h_{\beta \delta }\sigma
^{\gamma \delta }  \label{fourteen}
\end{equation}
where $\sigma _{\alpha \beta }$ is the shearing tensor 
\begin{equation}
\sigma _{\alpha \beta }=u_{\left( \alpha ;\beta \right) }+a_{(\alpha
}u_{\beta )}-\frac 13\Theta h_{\alpha \beta },  \label{fifteen}
\end{equation}
$\eta $ is some coefficient and $a_\alpha $ is the acceleration 
\begin{equation}
a_\alpha =u_{\alpha ;\beta }u^\beta .  \label{sixteen}
\end{equation}
The shearing tensor satisfies the conditions 
\begin{equation}
\sigma _{\alpha \beta }u^\beta =0,\quad \sigma _{\alpha \beta }g^{\alpha
\beta }=0,  \label{seventeen}
\end{equation}
hence, Eq (14) transforms into 
\begin{equation}
T_{\alpha \beta }^S=-2\eta \sigma _{\alpha \beta }.  \label{eighteen}
\end{equation}
Use of Eqs (5,15) gives the non-zero components of the shear \cite{four} 
\begin{equation}
\sigma _{11}=\frac 23B^2\sigma ,\quad \sigma _{22}=\frac{\sigma _{33}}{\sin
^2\theta }=-\frac 13R^2\sigma  \label{nineteen}
\end{equation}
where 
\begin{equation}
\frac 23\sigma ^2=\sigma ^{\alpha \beta }\sigma _{\alpha \beta }.
\label{twenty}
\end{equation}
One can check that the same components follow when $\sigma _{\alpha \beta }$
is written as the tensor 
\begin{equation}
\sigma _{\alpha \beta }=-\frac 13\sigma h_{\alpha \beta }+\sigma \chi
_\alpha \chi _\beta .  \label{twentyone}
\end{equation}
Thus it coincides with the general shear tensor defined by Eq (15) in the
spherically symmetric case. It also satisfies relations (17) in any metric.

Plugging Eq (21) into Eq (18) and comparing it to Eq (3) we find the
effectively generated pressures 
\begin{equation}
p_r^S=-2p_t^S=-\frac 43\eta \sigma .  \label{twentytwo}
\end{equation}
The degree of anisotropy is changed. There is no generation of energy
density or heat flow. The scalars $\Theta ,\sigma $ can be expressed through
the metric and its first derivatives: 
\begin{equation}
\Theta =\frac 1A\left( \frac{\dot B}B+2\frac{\dot R}R\right) ,\quad \sigma
=\frac 1A\left( \frac{\dot B}B-\frac{\dot R}R\right) .  \label{twthree}
\end{equation}

\section{Electromagnetic fields}

The EMT of electromagnetic fields is given by 
\begin{equation}
T_{\alpha \beta }^{EM}=\frac 1{4\pi }\left( F_{\mu \alpha }F_{\;\beta }^\mu
-\frac 14g_{\alpha \beta }F_{\mu \nu }F^{\mu \nu }\right)  \label{twfour}
\end{equation}
where $F_{\mu \nu }$ is the Faraday tensor. One defines a unit timelike
vector field $n^\mu $. An observer moving in its direction will measure
electric and magnetic field respectively 
\begin{equation}
E_\alpha =F_{\alpha \mu }n^\mu ,\quad H_\alpha =\frac 12\varepsilon _{\alpha
\mu \nu }F^{\mu \nu }.  \label{twfive}
\end{equation}
These fields are spacelike, $E^\alpha n_\alpha =H^\alpha n_\alpha =0$. The
Faraday tensor decomposes like \cite{ten} 
\begin{equation}
F_{\alpha \beta }=\varepsilon _{\alpha \beta \mu }H^\mu -2E_{[\alpha
}n_{\beta ]}.  \label{twsix}
\end{equation}
Plugging this expression into Eq (24) gives formula (7) from Ref \cite{ten},
which becomes after some rearrangements 
\begin{equation}
T_{\alpha \beta }^{EM}=\frac 1{4\pi }\left( H^2+E^2\right) \left( n_\alpha
n_\beta +\frac 12g_{\alpha \beta }\right) -\frac 1{4\pi }\left( E_\alpha
E_\beta +H_\alpha H_\beta \right) +2j_{(\alpha }n_{\beta )}  \label{twseven}
\end{equation}
where $E^2=E_\mu E^\mu ,H^2=H_\mu H^\mu $ and $j_\alpha $ is the Poynting
vector that measures the energy flow in the spacetime 
\begin{equation}
j_\alpha =\frac 1{4\pi }\varepsilon _{\alpha \mu \nu }E^\mu H^\nu .
\label{tweight}
\end{equation}

In order to absorb this EMT by the EMT for anisotropic fluid we choose the
direction $n^\alpha =u^\alpha $ and $\chi ^\alpha =E^\alpha /E$. The latter
is possible because when spherical symmetry is imposed $H_\alpha =0$ and $%
E_\alpha $ has only a radial spatial component. The would be heat flow term
in Eq (27) disappears and we get 
\begin{equation}
T_{\alpha \beta }^E=2eu_\alpha u_\beta +eg_{\alpha \beta }-2e\chi _\alpha
\chi _\beta ,\quad e=\frac{E^2}{8\pi }.  \label{twnine}
\end{equation}
Thus the addition of electric field induces effective pressures and energy
density 
\begin{equation}
\mu ^E=p_t^E=-p_r^E=e,  \label{thirty}
\end{equation}
related by simple linear equations of state. Hence, a charged perfect fluid
or a charged anisotropic fluid may be represented effectively by some
neutral anisotropic fluid. There is no heat flow induction in this case.

Charged fluids, however, must satisfy the Maxwell equations in addition to
the Einstein ones. It is clear from the above that the electromagnetic
tensor has a single component. The first pair of Maxwell equations gives $%
F_{10}=\Phi ^{\prime }$, where $\Phi $ is the only component of the
electromagnetic potential. The second pair of equations reads 
\begin{equation}
4\pi \tau u^\alpha =F_{\quad ;\beta }^{\alpha \beta }\equiv \left( -g\right)
^{-1/2}\left[ \left( -g\right) ^{1/2}F^{\alpha \beta }\right] _{,\beta }
\label{thone}
\end{equation}
where $\tau $ is the charge density, $g$ is the determinant of the metric,
usual derivative is denoted by comma and the covariant derivative is denoted
by semi-colon. In the spherically symmetric case this formula provides two
equations. One of them gives 
\begin{equation}
\frac{R^2}{AB}\Phi ^{\prime }=P\left( r\right) ,  \label{thtwo}
\end{equation}
$P\left( r\right) $ being an arbitrary function of the radius. Hence, the
combination in the l.h.s. depends only on the radial coordinate. Then the
second equation yields 
\begin{equation}
4\pi \tau =\frac{P^{\prime }}{BR^2}  \label{ththree}
\end{equation}
and becomes a formula for the charge density. Plugged in the definition of $E
$ the Maxwell equations lead to 
\begin{equation}
E=\frac{P\left( r\right) ^{\prime }}{R^2}  \label{thfour}
\end{equation}
which represents a constraint on the form of the effective density and
pressures $e$ in the general time-dependent case and no constraint in the
static case.

\section{Null fluid}

Null fluid describes dissipation in the free streaming approximation \cite
{four} and adds to the basic EMT the piece 
\begin{equation}
T_{\alpha \beta }^N=\varepsilon l_\alpha l_\beta  \label{thfive}
\end{equation}
where $l^\alpha $ is the null vector 
\begin{equation}
l^\alpha =A^{-1}\delta _0^\alpha +B^{-1}\delta _1^\alpha =u^\alpha +\chi
^\alpha ,  \label{thsix}
\end{equation}
satisfying the relations 
\begin{equation}
l^\mu l_\mu =0,\quad l^\mu u_\mu =-1.  \label{thseven}
\end{equation}
Substituting Eq (36) into Eq (35) one finds 
\begin{equation}
T_{\alpha \beta }^N=\varepsilon u_\alpha u_\beta +\varepsilon \chi _\alpha
\chi _\beta +\varepsilon \left( u_\alpha \chi _\beta +u_\beta \chi _\alpha
\right) .  \label{theight}
\end{equation}
A comparison of this expression with Eq (3), taking into account that $%
q^\alpha =q\chi ^\alpha $, shows that the addition of null fluid generates
effective energy density, radial pressure and heat flow, all of them equal 
\begin{equation}
\mu ^N=p_r^N=q^N=\varepsilon .  \label{thnine}
\end{equation}
No tangential pressure is generated. This is the only case where an
effective heat flow is induced.

\section{Summary}

The results in the previous sections show that viscosity, electric charge
and null fluids are equivalent to induced energy density and pressures,
related by simple linear equations of state $\mu =np_r$, $n=0,\pm 1$ and $%
p_t=kp_r$, $k=0,\pm 1,-1/2$. When all such additions are combined and
absorbed by the initial anisotropic fluid model, one obtains an effective
model with 
\begin{equation}
\mu ^e=\mu +e+\varepsilon ,  \label{forty}
\end{equation}
\begin{equation}
p_r^e=p_r-\zeta \Theta -\frac 43\eta \sigma -e+\varepsilon ,  \label{foone}
\end{equation}
\begin{equation}
p_t^e=p_t-\zeta \Theta +\frac 23\eta \sigma +e+\varepsilon ,  \label{fotwo}
\end{equation}
\begin{equation}
q^e=q+\varepsilon .  \label{fothree}
\end{equation}
These should be plugged into Eq (6) and hence in the l.h.s. of the Einstein
equations (7-10). There are 4 equations for 8 functions ($\mu
,p_r,p_t,q,\zeta ,\eta ,e,\varepsilon $). The quantities $\sigma ,\Theta $
are given by Eq (23). One has to impose 4 relations on these functions or
set some of them to zero in order to obtain a determined system of
equations. Several conclusions can be drawn.

Only the functions giving the two modes of dissipation of energy ($%
q,\varepsilon $) have effect upon the heat flow.

Viscosity ($\zeta ,\eta $) does not induce effective energy density.

An important characteristic of the fluid model is the anisotropy factor 
\begin{equation}
\bigtriangleup p^e=p_r^e-p_t^e=\bigtriangleup p-2\eta \sigma -2e.
\label{fofour}
\end{equation}
We see that shear viscosity and charge induce pressure anisotropy. Their
absorption by a perfect fluid ($\bigtriangleup p=0$) makes the latter
anisotropic and adds several more sources of anisotropy to the usual ones 
\cite{one}.

Bulk viscosity and null fluid don't induce anisotropy and may be absorbed by
an effective perfect fluid model with heat flow.

Charged perfect fluids are equivalent to anisotropic neutral fluids because 
\begin{equation}
\bigtriangleup p=-2e.  \label{fofive}
\end{equation}
In addition to the Einstein equations charged fluids satisfy Maxwell
equations but we have clarified that they give a formula for the charge
density and a mild constraint on $e$. Thus results about anisotropic fluids
may be carried over to charged perfect fluids and vice versa.

\section{The static case}

In this case there is no time dependence and $R=r,q=0$. The other three
Einstein equations reduce to 
\begin{equation}
8\pi \mu =\frac 1{r^2}-\frac{B^{-2}}r\left( \frac 1r-2\frac{B^{\prime }}%
B\right) ,  \label{fosix}
\end{equation}
\begin{equation}
8\pi p_r=-\frac 1{r^2}+\frac{B^{-2}}r\left( \frac 1r+2\frac{A^{\prime }}%
A\right) ,  \label{foseven}
\end{equation}
\begin{equation}
8\pi p_t=B^{-2}\left[ \frac{A^{\prime \prime }}A-\frac{A^{\prime }B^{\prime }%
}{AB}+\frac 1r\left( \frac{A^{\prime }}A-\frac{B^{\prime }}B\right) \right] .
\label{foeight}
\end{equation}
The difference of the last two equations gives a linear equation for $B^{-2}$%
. Introducing the variable $z$%
\begin{equation}
A^2=e^{2\int \left( z-2/r\right) dr}  \label{fonine}
\end{equation}
and solving the linear equation one obtains \cite{eleven} 
\begin{equation}
B^2=-\frac{z^2K^2}{r^6\left[ 2\int \frac z{r^8}\left( 1+8\pi \Delta
pr^2\right) K^2dr+C\right] }  \label{fifty}
\end{equation}
where $C$ is a constant of integration and 
\begin{equation}
K=e^{\int \left( \frac 2{zr^2}+z\right) dr}.  \label{fione}
\end{equation}
Thus the two generating functions $\Delta p$ and $z$ determine the metric of
any anisotropic fluid solution. Eqs (46-48) then determine its energy
density and the two pressures. In particular, using the results in the
previous sections, we can plug formulas (40-42) into the left hand sides of
Eqs (46-48) to obtain the general solution of any fluid with anisotropy,
shear and bulk viscosity and electric charge.

\section{Conclusions}

Anisotropic fluid models were studied for the first time by Lemaitre in 1933
on Einstein's recommendation (see the Golden oldie \cite{twelve} and the
references therein). His work, however, had no impact for a long time and
perfect fluid models have been examined for decades. In later times physical
reasons were given for anisotropy of pressures in relativistic star models 
\cite{one}, \cite{three} and now there exists an extensive literature on
this topic. In the present paper we have given additional arguments in
favour of the anisotropic fluid model with heat flow as an effective model,
encompassing fluids with anisotropy, viscosity, charge and radiation. We
have also given a vocabulary of the characteristics of different fluid
models in terms of the basic anisotropic one. Hence, one can concentrate on
finding its solutions and then translating them to the other models. This
procedure should yield numerous, albeit formal, relations between the
solutions of anisotropic and charged perfect models, or between viscous and
anisotropic models and so on. Such direction of research deserves a lot of
further study.


\begin{thebibliography}{99}
\bibitem{one}  Herrera, L., Santos, N.O.: Phys. Rep. \textbf{286}, 53 (1997).

\bibitem{two}  Ivanov, B.V.: Phys. Rev. D \textbf{65}, 104011 (2002).

\bibitem{three}  Bonnor, W.B., de Oliveira, A.K.G., Santos, N.O.: Phys. Rep. 
\textbf{181}, 269 (1989).

\bibitem{four}  Herrera, L., Santos, N.O., Wang, A.: Phys. Rev. D \textbf{78}%
, 084026 (2008).

\bibitem{five}  Chan, R., Herrera, L., Santos, N.O.: Mon. Not. R. Astron.
Soc. \textbf{267}, 637 (1994).

\bibitem{six}  Chan, R.: Mon. Not. R. Astron. Soc. \textbf{316}, 588 (2000).

\bibitem{seven}  Pinheiro, G., Chan, R.: Gen. Rel. Grav. \textbf{40}, 2149
(2008).

\bibitem{eight}  Ivanov, B.V.: Phys. Rev. D \textbf{65}, 104001 (2002).

\bibitem{nine}  Letelier, P.S.: Phys. Rev. D \textbf{22}, 807 (1980).

\bibitem{ten}  Lasky, P.D., Lun, A.W.C.: Phys. Rev. D \textbf{75}, 104010
(2007).

\bibitem{eleven}  Herrera, L., Ospino, J., Di Prisco, A.: Phys. Rev. D 
\textbf{77}, 027502 (2008).

\bibitem{twelve}  Lemaitre, G.: Gen. Rel. Grav. \textbf{29, }641 (1997).
\end{thebibliography}
\end{document}